\documentclass[twocolumn,preprintnumbers,amsmath,amssymb]{revtex4-1}
\usepackage{mathrsfs}
\usepackage{amssymb}
\usepackage{graphicx}
\usepackage{dcolumn}
\usepackage{bm}
\usepackage{textcomp}
\usepackage{amsmath}
\usepackage[titletoc]{appendix}
\usepackage{color}
\usepackage[modulo,pagewise]{lineno}
\usepackage{epstopdf}
\usepackage{fancyhdr}
\begin{document}
\title {Guessing probability in quantum key distribution}
\author{Xiang-Bin Wang$ ^{1,2,3*}$, Jing-Tao Wang$^{1*}$, Ji-Qian Qin$^1$, Cong Jiang$^1$, Zong-Wen Yu$^{1,4*}$}

\affiliation{ \centerline{$^{1}$State Key Laboratory of Low Dimensional Quantum Physics, Department of Physics,}
\centerline{Tsinghua University, Beijing 100084, China}
\centerline{$^{2}$Jinan Institute of Quantum technology, SAICT, Jinan 250101, China}
\centerline{$^{3}$Shenzhen Institute for Quantum Science and Engineering, and Physics Department,}
 \centerline{Southern University of Science and Technology, 518055 Shenzhen, China.}
\centerline{$^{4}$Data Communication Science and Technology Research Institute, Beijing 100191, China}
\centerline{$^{*}$email: xbwang@mail.tsinghua.edu.cn; jingtao0621@mail.tsinghua.edu.cn; yuzongwen@yeah.net}}
\begin{abstract}
\noindent On the basis of the existing trace distance result, we present a simple and efficient method to tighten the upper bound of the guessing probability. The guessing probability of the final key $\mathbf{k}$ can be upper bounded by the guessing probability of another key $\mathbf{k}^\prime$, if $\mathbf{k}^\prime$ can be mapped from the final key $\mathbf{k}$. Compared with the known methods, our result is more tightened by thousands of orders of magnitude. For example, given a $10^{-9}$-secure key from the sifted key, the upper bound of the guessing probability obtained using our method is $2\times10^{-3277}$. This value is smaller than the existing result $10^{-9}$ by more than $3000$ orders of magnitude. Our result shows that from the perspective of guessing probability, the performance of the existing trace distance security is actually much better than what was assumed in the past.
\end{abstract}

\maketitle
\section {Introduction}

\noindent The first quantum key distribution (QKD) protocol has been proposed by Bennett and Brassard in 1984; the protocol was based on the fundamentals of quantum mechanics~\cite{bennett1984quantum}. Since then, the security of QKD has always been the central issue in the quantum cryptographic field~\cite{renner2008security}. Trace distance is a very important security criterion ~\cite{curty2014finite,tomamichel2012tight}. It provides the universal composable security~\cite{ben2005universal,renner2005universally}, which can guarantee the security of key regardless of its application such as one-time pad (OTP). This is why many studies choose trace distance for the security criterion~\cite{konig2007small,tomamichel2012tight,curty2014finite,hayashi2012concise}.

In a classical practical cryptosystem, the impact of guessing probability on security is very important~\cite{alimomeni2012guessing,issa2017measuring}. Specifically, the key generated by the QKD protocol is not based on the presumed hardness of mathematical problems; thus, the eavesdropper Eve can only guess the final key via the measurement result of her probe. The guessing probability intuitively describes the probability that Eve can correctly guesses the final key, which can reflect the number of guesses that Eve requires to obtain the final key.

There are few studies on the guessing probability of QKD. Because there are more rigorous security criterions, such as the trace distance~\cite{ben2005universal,renner2005universally}, which gives the composable security. This makes the theoretical foundation for security of QKD crucially important. However, in the real application of QKD projects, customers often ask the question of guessing probability. The existing prior art results cannot give them a satisfactory upper bound~\cite{portmann2014cryptographic}. Consequently, some people questioned the security of QKD by relying on the prior art results of guessing probability~\cite{yuen2016security}. For example, according to the existing result~\cite{portmann2014cryptographic}, the guessing probability of the $\varepsilon$-secure key is approximately $10^{-9}$ if $\varepsilon$ is approximately $10^{-9}$.  From the perspective of guessing probability, the security of the value $ 10^{-9} $ is equivalent to that of a $ 30 $ perfect bits. The existing classical computer systems can easily crack such key. In practice, it is not unusual to request a much smaller guessing probability such as $10^{-100}$ or $10^{-1000}$. Therefore, it is beneficial to find a more tightened upper bound of guessing probability.

As an important criterion in cryptography, guessing probability alone cannot guarantee the security of the final key. However, the large value of the loose upper bound of the guessing probability does not indicate the insecurity of the final key~\cite{yuen2016security} because the value is not achievable by Eve, and one can find a more tightened value for the upper bound of the guessing probability. Here, by applying the trace distance criterion~\cite{renner2008security}, we find such tightened bound. We show that the guessing probability is actually smaller than the existing bound values by many orders of magnitude if one takes the privacy amplification by Toeplitz matrix. This shows that the trace distance criterion~\cite{renner2008security} can actually produce a much better result than what was assumed previously in the viewpoint of guessing probability.

\section{Results}

We consider the security definitions of a practical QKD protocol with finite size under the framework of composable security~\cite{canetti2001universally,muller2009composability,tomamichel2012tight,curty2014finite}. Suppose that Alice and Bob get two $N$-bit sifted key strings, $\mathbf{s}$ and $\mathbf{s}^\prime$. By performing an error correction and private amplification scheme, Alice gets a $n_1$-bit key $\mathbf{k}$, and Bob gets an estimate key $\mathbf{\hat{k}}$ of $\mathbf{k}$ from $\mathbf{s}$ and $\mathbf{s}^\prime$. The protocol is $\varepsilon_{\mathrm{cor}}$-correct if $P[\mathbf{k}\neq \mathbf{\hat{k}}]\le \varepsilon_{\mathrm{cor}}$. In general, the key $\mathbf{k}$ of Alice can be correlated with an eavesdropper system, and the density matrix of Alice and Eve is $\rho_{\mathrm{AE}}$. The protocol outputs an $\varepsilon$-secure key~\cite{konig2007small}, if
\begin{equation}\label{eq2}
\frac{1}{2}\parallel\rho_{\mathrm{AE}}-\rho_{\mathrm{U}}\otimes\rho_{\mathrm{E}}\parallel_1\le \varepsilon,
\end{equation}
where $\parallel \centerdot \parallel_1$ denotes the trace norm, $\rho_{\mathrm{U}}$ is the fully mixed state of Alice's system. The protocol is $\varepsilon_{\mathrm{tol}}$-secure if $\varepsilon_{\mathrm{cor}}$ and $\varepsilon$ satisfy $\varepsilon_{\mathrm{cor}}+\varepsilon\le \varepsilon_{\mathrm{tol}}$, which means that it is $ \varepsilon_{\mathrm{tol}} $-indistinguishable from a perfect protocol (which is correct and secret). Without any loss of generality, we consider the case of $ \varepsilon_{\mathrm{cor}}=\varepsilon $ in this article.

We define the security level: \\

\newtheorem{law}{Definition}
\begin{law}
	
If key $\mathbf{k}$ is $\varepsilon$-secure, the \textbf{security level} of key $\mathbf{k}$ is $\varepsilon$.

\end{law}
For symbol clarity, we will use notation $\varepsilon_{\mathbf{k}}$ for the security level of key $\mathbf{k}$. With this definition, we can say that the key $\mathbf{k}$ is $\varepsilon_{\mathbf{k}}$-secure or that its security level is $\varepsilon_{\mathbf{k}}$.

We define the guessing probability:\\
\begin{law}
	 Let the final key generated by the QKD protocol be $\mathbf{k}$; the \textbf {guessing probability} of $\mathbf{k}$ is defined as the success probability of the attacker Eve guessing the final key via her measurement result and is denoted as $p(\mathbf{k})$.
\end{law}

\newtheorem{lemm}{Lemma}
\begin{lemm}
	The guessing probability of $\varepsilon_{\mathbf{k}}$-secure key $\mathbf{k}$ with length $n_1$ is not larger than $\frac{1}{2^{n_1}} + \varepsilon_{\mathbf{k}}$.
\end{lemm}

This is a conclusion from Ref.~\cite{portmann2014cryptographic}. The proof has been already given in Ref.~\cite{portmann2014cryptographic}; for the convenience of readers, we write the proof again in the Method section.

According to Lemma 1, the guessing probability of key $\mathbf{k}$ can be divided into two parts; one part $ 2^{-n_1} $ is related to the length of the key, the other part $ \varepsilon_{\mathbf{k}}(n_1) $ is related to the security level. Under the framework of universally composable security, when calculating the final key length, we often make the security level to be between $10^{-9} \sim 10^{-24}$, which is much bigger than $2^{-n_1}$  because $n_1$ is often $ 10^3, 10^4$, or larger. Therefore, $2^{-n_1}$ can be ignored and $p(\mathbf{k})\leq \bar{p}(\mathbf{k}) \sim \mathcal{O}(\varepsilon(\mathbf{k}))$. However, the guessing probability of a secure key with a length of tens of bits can also reach this magnitude. Therefore, when the secure requirements are very high, it is clearly not enough for a key with a length of thousands of bits or even longer if the upper bound of guessing probability only stops at this magnitude. Therefore, we cannot simply use this formula alone to obtain the upper bound of the guessing probability. Fortunately, we have a much better way for tightening the bound. The approach will be presented below.

\begin{lemm}
If key $\mathbf{k}$ can be mapped to string $\mathbf{k}^\prime$ by a map $M$ that is known to Eve, then the guessing probability of $\mathbf{k}$ cannot be larger than the guessing probability of string $\mathbf{k}^\prime$, i.e.,
\begin{equation}
p (\mathbf{k})\le p (\mathbf{k}^\prime).
\end{equation}
Here $p(\mathbf{k}), p(\mathbf{k}^\prime)$ are the guessing probabilities of $\mathbf{k}$ and $\mathbf{k}^\prime$, respectively.
\end{lemm}
{\it Proof.} This lemma is clear because when Eve can correctly guess $\mathbf{k}$, Eve can obtain $\mathbf{k}^\prime$ by knowing the map $M$. Otherwise, Eve can still correctly guess the $\mathbf{k}^\prime$ with a probability not less than $0$, i.e., $p(\mathbf{k}^\prime) = p(\mathbf{k}) + \delta, \delta \ge 0$.


\newtheorem{theor}{Theorem}
\begin{theor}\label{theor1}
If the $\varepsilon_{\mathbf{k}}$-secure key $\mathbf{k}$ with a length $n_1$ can be mapped to the $\varepsilon_{\mathbf{k}^\prime}$-secure key $\mathbf{k}^\prime$ with length $n_2$, the guessing probability of $\mathbf{k}$ cannot be larger than $\mathbf{k}^\prime$, i.e.,
\begin{equation}\label{th1}
p(\mathbf{k}) \le \bar p(\mathbf{k}^\prime)= \frac{1}{2^{n_2}} + \varepsilon_{\mathbf{k}^\prime}.
\end{equation}
\end{theor}

\textit{Proof.} This theorem actually requires two conditions:

\noindent i) the final key $\mathbf{k}$ can be mapped to the string $\mathbf{k}^\prime$,

\noindent ii) the string $\mathbf{k}^\prime$ can be regarded as a $\varepsilon_{\mathbf{k}^\prime}$-secure key.

Using the above-mentioned conditions, the proof is very simple. Given the condition i), we can apply Lemma 2 to obtain
\begin{equation}\label{ee1}
p (\mathbf{k})\le p (\mathbf{k}^\prime).
\end{equation}
Given the condition ii), we can apply Lemma 1 to obtain
\begin{equation}\label{ee2}
p(\mathbf{k}^\prime) \leq \bar{p}(\mathbf{k}^\prime)=\frac{1}{2^{n_2}} + \varepsilon_{\mathbf{k}^\prime},
\end{equation}
where $\bar p(\mathbf{k}^\prime)$ is the upper bound of $p(\mathbf{k}^\prime)$. According to Eqs.~\eqref{ee1} and~\eqref{ee2}, we can obtain
\begin{equation}
p(\mathbf{k}) \le \bar p(\mathbf{k}^\prime)= \frac{1}{2^{n_2}}  + \varepsilon_{\mathbf{k}^\prime}.
\end{equation}
This ends our proof of Theorem 1.

As discussed above, if the length of the final key $\mathbf{k}$ and the string $\mathbf{k}^{\prime}$ are very large, then $2^{-n_1}$ and $2^{-n_2}$ can be ignored. Meanwhile, if $n_2<n_1$ and $\varepsilon_{\mathbf{k}^{\prime}}<\varepsilon_{\mathbf{k}}$, then $2^{-n_2}+\varepsilon_{\mathbf{k}^{\prime}}\sim \varepsilon_{\mathbf{k}^{\prime}} \leq \varepsilon_{\mathbf{k}} \sim 2^{-n_1}+\varepsilon_{\mathbf{k}}$. Thus, Theorem 1 can provide a tighter upper bound of guessing probability.

Using Theorem 1, it is now possible for us to obtain the upper bound of the guessing probability of the $\varepsilon_\mathbf{k}$-secure key $\mathbf{k}$ more tightly. Instead of directly applying Lemma 1, we choose to first map $k$ to a $n_2$-bit string $\mathbf{k}^\prime = M(\mathbf{k})$. If the string $\mathbf{k}^\prime$ itself can be regarded as an $\varepsilon_{\mathbf{k}^\prime}$-secure final key, we can apply Theorem 1 by calculating $\bar p(\mathbf{k}^\prime)$. In addition, we can obtain a much smaller upper bound of the guessing probability of $\mathbf{k}$ if $\varepsilon_{\mathbf{k}^\prime}$ is very small and $n_2$ is not too small. Now, the remaining problems are to determine the map $M$, to make sure that $\mathbf{k}^\prime=M(\mathbf{k})$ is another key that is $\varepsilon_{\mathbf{k}^\prime}$-secure, and to calculate $\varepsilon_{\mathbf{k}^{\prime}}$. We start our method with the hashing function in the key distillation.

\textbf{Our hashing function.} We use the key distillation with the random matrix. Denote $R_{nN}$ as the $n \times N$ random matrix with each element being randomly chosen to be either $0$ or $1$. In addition, we represent the $N$-bit sifted string $\mathbf{s}$ by a column vector, which contains $N$ elements. To obtain the $n$-bit final key, we use the calculation $R_{nN}\mathbf{s}$. It can be easily confirmed that our random matrix belongs to the class of two-universal hashing function family\cite{renner2008security}.

Suppose we have distilled out the $n_1$-bit key $\mathbf{k}$ from the $N$-bit sifted key $\mathbf{s}$ through hashing by our random matrix $R_{n_1N}$. We can map the $n_1$-bit key $\mathbf{k}$ into the $n_2$-bit string $\mathbf{k}^\prime=M(\mathbf{k})$ by deleting the last $n_1-n_2$ bits from the key string $\mathbf{k}$. Clearly, this string $\mathbf{k}^\prime$ mapped from $\mathbf{k}$ can be also regarded as another final key distilled from the sift key $\mathbf{s}$ by the $n_2 \times N$ random hashing matrix $R_{n_2N}$, which is a submatrix of  $R_{n_1N}$. In summary, we have
\begin{equation}
\mathbf{k}^\prime = M(\mathbf{k})=  R_{n_2N}\mathbf{s}.
\end{equation}

This means that $\mathbf{k}^\prime$ is a string mapped from key $\mathbf{k}$. Moreover, $\mathbf{k}^\prime$ can be regarded as another final key of length $n_2$ distilled from the sifted key $\mathbf{s}$. Because the two conditions in Theorem 1 are satisfied, according to Theorem 1, we can obtain a tightened upper bound of $p(\mathbf{k})$ with Eq.~\eqref{th1} if we know the security level of key $\mathbf{k}^\prime$, i.e., the value of $\varepsilon_{\mathbf{k}^\prime}$. Because our random matrix is a class of two-universal hashing function, the value $\varepsilon_{\mathbf{k}^\prime}$ depends on $n_2$~\cite{tomamichel2012tight}. The details are shown in the Method section and explain the calculation of $\varepsilon_{\mathbf{k}^\prime}$ for $n_2$. Hence, in the QKD protocol that uses a random hashing matrix presented here, to obtain the upper bound of the guessing probability of the $n_1$-bit final key $\mathbf{k}$, we can summarize the procedure above by the following scheme:

\textbf{Scheme}
1) Given the $n_1$-bit final key $\mathbf{k}$, we delete its last $n_1-n_2$ bits and obtain a string $\mathbf{k}^\prime$. 2) We regard $\mathbf{k}^\prime$ as another possible final key that is $\varepsilon_{\mathbf{k}^\prime}$-secure. Compute the $\varepsilon_{\mathbf{k}^\prime}$ value of $\mathbf{k}^\prime$ with the input parameters $N$ and $n_2$. 3) Calculate $\bar p(\mathbf{k})$ by Theorem 1 through Eq.~\eqref{th1}.

Because in our scheme the value of $\varepsilon_{\mathbf{k}^\prime}$ is dependent on $n_2$, as shown in the Method section, we can now replace $\varepsilon_{\mathbf{k}^\prime}$ by a functional form, $\varepsilon_{\mathbf{k}^\prime}(n_2)$. To obtain the tightened upper bound value of the guessing probability in scheme 1,  we need to choose an appropriate $n_2$ value. In our calculation, we set the condition
\begin{equation}\label{ee3}
2^{-n_2} = \varepsilon_{\mathbf{k}^\prime}(n_2),
\end{equation}
for the appropriate $n_2$.

For any $n > n_2$, we have $\varepsilon_{\mathbf{k}}(n)>\varepsilon_{\mathbf{k}^{\prime}}(n_2)=2^{-n_2}$; however, for any $n < n_2$, we have $2^{-n}> 2^{-n_2}$. In conclusion, if $n \neq n_2$, $2^{-n}+ \varepsilon_{\mathbf{k}}(n)> 2^{-n_2}$. Therefore, in this study, we set $2^{-n_2} = \varepsilon_{\mathbf{k}^\prime}(n_2)$, and obtain a tightened guessing probability $2^{-n_2+1}$.

Once we determine the value $n_2$ and the corresponding $\varepsilon_{\mathbf{k}^\prime}(n_2)$, we calculate $\bar p(\mathbf{k}^\prime)$ by Eq.~\eqref{th1}. Clearly, this is the upper bound of the guessing probability of the final key $\mathbf{k}$ of length $n_1$ provided that
\begin{equation}
n_1 > n_2.
\end{equation}
Thus, we can actually use a more efficient scheme to obtain the upper bound of the guessing probability of key $\mathbf{k}$, as the following Theorem 2:

\begin{theor}\label{theor2}
	In the QKD protocol, if the $n_1$-bit final key $\mathbf{k}$ is distilled from the sifted key $\mathbf{s}$ using a random matrix $R_{n_1N}$, the guessing probability of $\mathbf{k}$ can be upper bounded by
	\begin{equation}
	p(\mathbf{k}) \le \bar p(\mathbf{k}^\prime) =2^{-(n_2-1)},
	\end{equation}
	where  $ \mathbf{k}^\prime = M(\mathbf{k})=  R_{n_2N}\mathbf{s} $ and $n_2$ satisfies $2^{-n_2}=\varepsilon_{\mathbf{k}^\prime}(n_2), n_2 < n_1.$
\end{theor}

\begin{figure}
	\centering
	\includegraphics[scale=0.4]{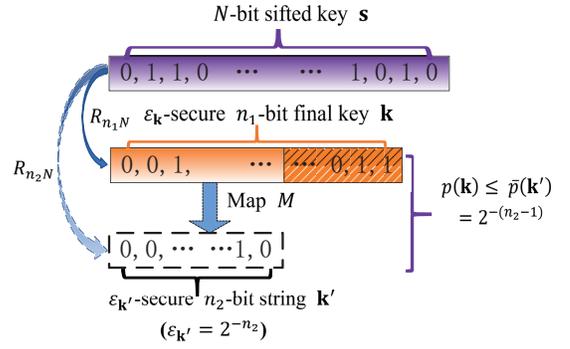}
	\caption{Flow chart of our method of bounding the guessing probability. The arrow between $\mathbf{s}$ and $\mathbf{k}$ indicates that the $\varepsilon_{\mathbf{k}}$-secure $n_1$-bit final key $\mathbf{k}$ can be distilled from the $N$-bit sifted key $\mathbf{s}$ using a random matrix $R_{n_1 N}$, i.e., $\mathbf{k}= R_{n_1 N}\mathbf{s}$. The arrow between $\mathbf{k}$ and $\mathbf{k}^{\prime}$ indicates that there exists a map $M$ that can map the key $\mathbf{k}$ into $\mathbf{k}^{\prime}$, i.e. $\mathbf{k}^{\prime}=M(\mathbf{k})$. The arrow between the sifted key $S$ and $\mathbf{k}^{\prime}$ indicates that if a random hashing matrix $R_{n_2 N}$ is used to distill the final key, we have $\mathbf{k}^{\prime}=R_{n_2 N}\mathbf{s}$. Then, if $n_2$ satisfies the condition in Theorem~\ref{theor2}, a tightened guessing probability of $\mathbf{k}$ can be obtained.}
	\label{fi1}
\end{figure}

 As shown in Fig.~\ref{fi1}, the arrow between $\mathbf{s}$ and $\mathbf{k}$ indicates that the $\varepsilon_{\mathbf{k}}$-secure $n_1$-bit final key $\mathbf{k}$ can be distilled from the $N$-bit sifted key $\mathbf{s}$ using a random matrix $R_{n_1 N}$, i.e. $\mathbf{k}= R_{n_1 N}{\mathbf{s}}$. The arrow between $\mathbf{k}$ and $\mathbf{k}^{\prime}$ indicates that there exists a map $M$ that can map the key $\mathbf{k}$ into $\mathbf{k}^{\prime}$, i.e., $\mathbf{k}^{\prime}=M(\mathbf{k})$. The arrow between the sifted key $\mathbf{s}$ and $\mathbf{k}^{\prime}$ indicates that if a random hashing matrix $R_{n_2 N}$ is used to distill the final key, we have $\mathbf{k}^{\prime}=R_{n_2 N}\mathbf{s}$. Then if $n_2$ satisfies the condition in Theorem~\ref{theor2}, a tightened guessing probability of $\mathbf{k}$ can be obtained.


There are two important points need to be noticed. First, when applying our theorem to obtain the non-trivial upper bound of the guessing probability for the final key $\mathbf{k}$, we do not really need to transform $\mathbf{k}$ to another string $\mathbf{k}^{\prime}$, and we only need the existence of a map that can map $\mathbf{k}$ to $\mathbf{k}^{\prime}$ mathematically. That is to say, we use the final key $\mathbf{k}$, but its guessing probability is calculated from the shorter key $\mathbf{k}^{\prime}$. As shown above, the existence has been proven. Second, in this study, we use the random matrix $R_{nN}$ as a family of two-universal hash functions to distill the key to illustrate our conclusion more intuitively. Of course, we can also use the modified Toeplitz matrix~\cite{hayashi2012concise} instead of the random matrix $R_{nN}$. Thus, the final key $\mathbf{k}$ can be also mapped to the string $\mathbf{k}^{\prime}$, and the string $\mathbf{k}^{\prime}$ can also be regarded as the $\varepsilon_{\mathbf{k}^{\prime}}$-secure key. This means that the proposed theorem in this study still holds.

\section{Discussion}

\begin{table}[]
	
	\vspace{20pt}
	\centering
	\begin{tabular}{p{2cm}p{2cm}p{2cm}p{2cm}p{2cm}}
		\hline
		\hline
		$ N_{\mathrm{tol}} $ & $ 10^4 $ & $ 10^5 $ & $ 10^6 $  \\
		\hline
		$ n $  & $ 2.01\times10^3 $   & $ 4.06\times10^4 $   & $ 4.90\times10^5 $       \\
		$ p_{\mathrm{g}} $\cite{yuen2016security}  & $ 10^{-6} $   & $ 10^{-6} $   & $ 10^{-6}$    \\
		$ p_{\mathrm{g}} $\cite{portmann2014cryptographic}  & $ 10^{-9} $  & $ 10^{-9} $  & $ 10^{-9} $   \\
		$ p_{\mathrm{g}}^{\mathrm{Thm.2}} $  & $ 2\times10^{-32} $   &  $ 2\times10^{-327} $  &   $ 2\times 10^{-3277} $          \\
		\hline
		\hline
	\end{tabular}
	\caption{Comparison of the guessing probability, where $ Q_{\mathrm{tol}}=2.14\%$ is the channel error tolerance, $ N_z=0.22N_{\mathrm{tol}}$ is the length of the string used to do parameter estimation, $N_{\mathrm{tol}}$ is the total length of the sifted key, $N=0.78N_{\mathrm{tol}}$ is the length of the string for key generation, $ \varepsilon=10^{-9}$ is the security level, $n$ is the length of the $10^{-9}$-secure key, and $p_{\mathrm{g}}$ is the probability of correctly guessing the final key. Specifically, $ p_{\mathrm{g}}^{\mathrm{Thm.2}} $ is the result of Theorem 2 of this work.}
	\label{bs1}
\end{table}

\begin{table}[]
	
	\vspace{20pt}
	\centering
	\begin{tabular}{p{2cm}p{2cm}p{2cm}p{2cm}p{2cm}}
		\hline
		\hline
		$ N_{\mathrm{tol}} $ & $ 10^4 $ & $ 10^5 $ & $ 10^6 $  \\
		\hline
		$ \varepsilon $  & $ 10^{-9} $  & $ 10^{-9} $  & $ 10^{-9} $    \\
		$ n $  & $ 2.01\times10^3 $   & $ 4.06\times10^4 $   & $ 4.90\times10^5 $       \\
		$ r $  & $ 0.20 $ & $ 0.41 $   & $ 0.49 $       \\
		$ \varepsilon^\prime$  & $ 10^{-32} $   &  $ 10^{-327} $  &   $ 10^{-3277} $    \\
		$ n^\prime $  & $ 136 $ & $ 1.12\times10^3 $ & $ 1.10\times10^4 $  \\
		$ r^\prime $  & $ 0.01 $ & $ 0.01 $   & $ 0.01 $       \\
		\hline
		\hline
		\label{bs2}
	\end{tabular}
	\caption{Comparison of the rate $ r=n/N_{\mathrm{tol}} $ and $ r^\prime=n^\prime/N_{\mathrm{tol}} $ under the same parameters shows in Table.~\ref{bs1}. $ \varepsilon$ and $ \varepsilon^\prime$ are the security levels, $n$ and $n^\prime$ are the length of $\varepsilon$-secure key and the length of $\varepsilon^\prime$-secure key, respectively.}
	\label{bs2}
\end{table}
\newcommand{\RNum}[1]{\uppercase\expandafter{\romannumeral #1\relax}}

Table.~\ref{bs1} describes the upper bounds of the guessing probability calculated by different $N_{\mathrm{tol}}$, where $N_{\mathrm{tol}}$ is the length of the total string that includes the sifted keys for key generation and the string used to do parameter estimation. In Table.~\ref{bs1}, $ N_{\mathrm{tol}}=10^4,10^5$, and $10^6$. Table.~\ref{bs1} shows that when $ N_{\mathrm{tol}} = 10^6 $,  $ n=4.90\times10^5 $ and the guessing probabilities obtained using the methods of ~\cite{yuen2016security} and ~\cite{portmann2014cryptographic} are approximately $ 10^{-6} $ and $ 10^{-9} $. However, using our method, the guessing probability can be reduced to $ 2\times10^{-3277} $, which is more tightened by thousands of orders of magnitude than prior art methods. With an increase in the length of $N_{\mathrm{tol}}$, the length of the final key also increases; however, the guessing probabilities in~\cite{yuen2016security} and ~\cite{portmann2014cryptographic} almost remain unchanged. Compared with~\cite{yuen2016security} and~\cite{portmann2014cryptographic}, the guessing probability obtained by our method is considerably reduced, which is more realistic and tighter. It should be noted that we calculate the case without the known-plaintext attack (KPA) in Table.~\ref{bs1}. Now, we consider the case of KPA in QKD using our method. Suppose that Eve knows the $t$ bits of the final $n_2$-bit key $\mathbf{k}^\prime$; then, the guessing probability of the $\varepsilon_{\mathbf{k}^\prime}$-secure key $\mathbf{k}^\prime$ is $p_{\mathrm{KPA}}(\mathbf{k}^\prime)\leq 2^{-(n_2-t-1)}$. Now, the upper bound of the guessing probability of key $\mathbf{k}^\prime$ is equal to that of an ideal $(n_2-t-1)$-bit key.

Table.~\ref{bs2} compares the length of the $ \varepsilon $-secure key $ n $ and the length of $ \varepsilon^\prime $-secure key $ n^\prime $  when the total length of the sifted key is $ 10^4,10^5$, and $10^6 $. This table shows that if only using Lemma 1 to obtain a smaller guessing probability, $\varepsilon$ needs to be reduced. Accordingly, the length of the final key and the key rate will be considerably reduced. For example, from Table II, when $ N_{\mathrm{tol}}=10^6 $, if the customer wants to reduce the guessing probability from $10^{-9}$ to $2 \times 10^{-3277}$, the length of the key will become $ n^{\prime}=1.1\times10^4 $, and the key rate will become $ r^{\prime}=0.01 $. This result is much lower than the original key length $ n=4.9\times10^5 $ and the key rate $r= 0.49 $. Using our result, there is actually no bit cost for a much smaller bound value of guessing probability. For example, when $N_{\mathrm{tol}}=10^6$, we can upper bound the guessing probability by $2 \times 10^{-3277}$ by setting $\varepsilon = 10^{-9}$. Thus, without reducing the value of $\varepsilon$, we can obtain a tightened upper bound of guessing probability $ p_{\mathrm{g}}^{\mathrm{Thm.2}} $ of $ \mathbf{k} $, as can be seen from Table.~\ref{bs1}.

Our result shows that in terms of guessing probability, the performance of the existing trace distance security is much better than what has been assumed in the past. Incidentally, after the upper bound value $10^{-9}$ was presented in Ref.~\cite{portmann2014cryptographic}, a looser upper bound, $10^{-6}$ for guessing probability was presented~\cite{yuen2016security}. We emphasize that this looser upper bound does not in any sense challenge the validity of the existing security proof of QKD~\cite{portmann2014cryptographic}. Although the large value of {\it lower bound} of Eve's guessing probability can show insecurity, the large
value of {\it upper bound} cannot show insecurity. If one does not make any effort, one can also obtain a large-value upper bound of $100\%$ for Eve's guessing probability. Such value is correct for the upper bound but not meaningful. If any new upper bound is larger than that in the prior art result, it means that the "new upper bound" is trivial and meaningless rather than the prior art result is invalid. Thus, the looser upper bound presented by Ref.~\cite{yuen2016security} only shows that Eve's guessing probability of the key is smaller than $10^{-6}$. It does not conflict with more tightened results presented elsewhere.

In this study, our goal is to obtain a tightened guessing probability. On the basis of the existing secure criterion (Trace distance) and the general property of guessing probability, we propose a simple and efficient method to tighten the upper bound of the guessing probability. We find that the guessing probability $p(\mathbf{k})$ of $\mathbf{k}$ can be upper bounded by $2^{-(n_2-1)}$, where $n_2$ satisfies $2^{-n_2}=\varepsilon_{\mathbf{k}^\prime}(n_2)$ and $n_2 < n_1$. Specifically, a simple random matrix $R_{nN}$ can be used to distill the final key. Compared with the prior art results, of which the upper bound of the guessing probability of the $\varepsilon$-secure key is approximately $\varepsilon$, our method provides a more tightened upper bound. Therefore, the loose upper bound for the guessing probability obtained in Ref.~\cite{yuen2016security} cannot be regarded as evidence to question the validity of existing the security proof of QKD.

\section{Method}
\setcounter{lemm}{0}
\noindent{\bf Proof of Lemma 1}
\begin{lemm}
		The guessing probability of the $\varepsilon_{\mathbf{k}}$-secure key $\mathbf{k}$ with length $n_1$ is not larger than $\frac{1}{2^{n_1}} + \varepsilon_{\mathbf{k}}$.
\end{lemm}

This is a conclusion obtained from Ref.~\cite{portmann2014cryptographic}. The proof has been already presented in Ref.~\cite{portmann2014cryptographic}. Here, for the convenience of the reader, we write the proof again.

{\it Proof.} Let the $n$-bit string $\mathbf{x}$ be the $\varepsilon_{\mathbf{x}}$-secure key in $\mathcal{X}$. The density matrix of Alice and Eve is $\rho_{\mathrm{XE}}$ and satisfies

\begin{equation}\label{meq1}
\begin{aligned}
&\rho_{\mathrm{XE}}=\sum_{\mathbf{x}\in \mathcal{X}}|\mathbf{x}\rangle \langle \mathbf{x}|\otimes\rho_E^{\mathbf{x}},\\
&\frac{1}{2}\parallel\rho_{\mathrm{XE}}-\rho_{\mathrm{U}_{\mathbf{x}}}\otimes\rho_{\mathrm{E}}\parallel_1\le \varepsilon_{\mathbf{x}},
\end{aligned}
\end{equation}

where $\rho_{\mathrm{U}_{\mathbf{x}}}$ is the fully mixed state in $\mathcal{X}$. Then we have
\begin{equation}\label{meq2}
\begin{aligned}
& \frac{1}{2}\| \rho_{\mathrm{XE}}-\rho_{\mathrm{U}_{\mathbf{x}}}\otimes\rho_{\mathrm{E}}\|_1 \\
\ge & \frac{1}{2}\|\sum_{\mathbf{x} \in \mathcal{X}} q(\mathbf{x}) |\mathbf{x}\rangle \langle \mathbf{x}|-\sum_{\mathbf{x} \in \mathcal{X}} \frac{1}{2^{n}}|\mathbf{x}\rangle \langle \mathbf{x}|\|_1\\
= & \frac{1}{2}\sum_{\mathbf{x} \in \mathcal{X}}|q(\mathbf{x})-\frac{1}{2^{n}}|.\\
\end{aligned}
\end{equation}
Eve's guessing probability of string $\mathbf{x}$ is $q(\mathbf{x})$, and the maximum guessing probability is $p_{\mathrm{g}} = \max_{\mathbf{x} \in \mathcal{X}} \{q(\mathbf{x})\}$. Without any loss of generality, it is possible to assume that the maximum guessing probability is $q(\mathbf{x}^\prime)$. Note that $\sum_{\mathbf{x} \in \mathcal{X}} q(\mathbf{x}) =1$, then the following holds

\begin{equation}\label{meq3}
\begin{aligned}
& \frac{1}{2}\sum_{\mathbf{x} \in \mathcal{X}}|q(\mathbf{x})-\frac{1}{2^{n}}| \\
= & \frac{1}{2}|q(\mathbf{x}^\prime)-\frac{1}{2^{n}}|+\frac{1}{2}\sum_{\mathbf{x} \in \mathcal{X},\mathbf{x} \neq \mathbf{x}^\prime}|q(\mathbf{x})-\frac{1}{2^{n}}|\\
\geq & \frac{1}{2}|q(\mathbf{x}^\prime)-\frac{1}{2^{n}}|+\frac{1}{2}|\sum_{\mathbf{x} \in \mathcal{X}, \mathbf{x} \neq \mathbf{x}^\prime}[q(\mathbf{x})-\frac{1}{2^{n}}]|\\
= & |q(\mathbf{x}^\prime)-\frac{1}{2^{n}}|.\\
\end{aligned}
\end{equation}
From Eqs.\eqref{meq1}-\eqref{meq3}, we have $p_{\mathrm{g}}\le {2^{-n_1}} + \varepsilon_{\mathbf{x}}$; thus, for the $n_1$-bit $\varepsilon_{\mathbf{k}}$-secure key $\mathbf{k}$, the guessing probability satisfies
\begin{equation}\label{barpk}
p(\mathbf{k}) \leq \bar{p}(\mathbf{k})=\frac{1}{2^{n_1}} + \varepsilon_{\mathbf{k}},
\end{equation}
where $\bar p(\mathbf{k})$ is the upper bound of $p(\mathbf{k})$. This ends our proof of \textit{Lemma 1}.

\noindent{\bf Calculation of $\varepsilon_{\mathbf{k}^\prime}$}

We consider the security definitions of a practical QKD protocol with a finite-size under the framework of composable security~\cite{canetti2001universally,muller2009composability,tomamichel2012tight}. Suppose that Alice and Bob get two $N$-bit sifted key strings. By performing an error correction and private amplification scheme, Alice get a $n$-bit final key $\mathbf{k}$ and Bob get an estimate $\mathbf{\hat{k}}$ of $\mathbf{k}$. The protocol is $\varepsilon_{\mathrm{cor}}$-correct if $P[\mathbf{k}\neq \mathbf{\hat{k}}]\le \varepsilon_{\mathrm{cor}}$. In general, the key $\mathbf{k}$ of Alice can be correlated with an eavesdropper system, and the density matrix of Alice and Eve is $\rho_{\mathrm{AE}}$.

The protocol outputs an $\varepsilon_{\mathbf{k}}$-secure key~\cite{canetti2001universally}, if
\begin{equation}
\frac{1}{2}\parallel\rho_{\mathrm{AE}}-\rho_{\mathrm{U}}\otimes\rho_{\mathrm{E}}\parallel_1\le \varepsilon_{\mathbf{k}},
\end{equation}
where $\parallel \centerdot \parallel_1$ denotes the trace norm, $\rho_{\mathrm{U}}$ is the fully mixed state of Alice's system. The protocol is $\varepsilon_{\mathrm{tol}}$-secure if $\varepsilon_{\mathrm{cor}}$ and $\varepsilon_{\mathbf{k}}$ satisfies $\varepsilon_{\mathrm{cor}}+\varepsilon_{\mathbf{k}}\le \varepsilon_{\mathrm{tol}}$, which means that it is $ \varepsilon_{\mathrm{tol}} $-indistinguishable from an ideal protocol. Without any loss of generality, we consider the case of $ \varepsilon_{\mathrm{cor}}=\varepsilon_{\mathbf{k}}$.


From {\it Lemma 1}, we can calculate $ \bar p(\mathbf{k}) $ given the $n$-bit $\varepsilon_{\mathbf{k}}$-secure key $\mathbf{k}$. In this situation, $ \bar p(\mathbf{k}) =2^{-n}+\varepsilon_{\mathbf{k}}$. However, in our method, we only know $ N $ and  $ n_2 $, which are the length of the sifted key and  $\mathbf{k}^\prime$. (The string $\mathbf{k}^\prime$ itself can be also regarded as another final key distilled from the sifted key.) To get a tightened upper bound of the guessing probability of $ \mathbf{k} $, we need to obtain the value of $\varepsilon_{\mathbf{k}^\prime}$ .
According to Ref.~\cite{tomamichel2012tight}, with $N$ and $n_2$, the final key is $\varepsilon_{\mathbf{k}^\prime}$-secure if $\varepsilon_{\mathbf{k}^\prime}$ satisfies the following equation:

\begin{equation}\label{meq4}
n_2\leq N[1-h(Q_{\mathrm{tol}}+\mu)]-fNh(Q_{\mathrm{tol}})-\log\frac{2}{\varepsilon_{\mathbf{k}^\prime}^{3}}
\end{equation}
where  $ \mu =\sqrt{\frac{N+N_{z}}{NN_{z}} \frac{N_{z}+1}{N_{z}} \ln\frac{2}{\varepsilon_{\mathbf{k}^\prime}}} $, $ N_z$ is the length of string used for parameter estimation, $f=1.1$, $ h $ denotes the binary Shannon entropy function, $ h(x)=-x\log x-(1-x)\log(1-x) $ and $ Q_{\mathrm{tol}} $ represents the channel error tolerance. To obtain non-trivial results, we use equality in Eq.~\eqref{meq4} to calculate the value of $\varepsilon_{\mathbf{k}^\prime}$, given the input $n_2$. Since $\varepsilon_{\mathbf{k}^\prime}$ is dependent on $n_2$, we use notation $\varepsilon_{\mathbf{k}^\prime}(n_2)$ for $\varepsilon_{\mathbf{k}^\prime}$. Here, $\varepsilon_{\mathbf{k}^\prime}(n_2)$, if $n_2$ is given and we numerically find the value of $\varepsilon_{\mathbf{k}^\prime}$ by Eq.~\eqref{meq4}.

In our calculation, we choose specific a $n_2$-value that satisfies
\begin{equation}
2^{-n_2}=\varepsilon_{\mathbf{k}^\prime}(n_2).
\end{equation}
In combination with Eq.~\eqref{meq4},we obtain the following equation for the tightened $\varepsilon_{\mathbf{k}^\prime}$ value:

\begin{equation}\label{meq5}
-\log\varepsilon_{\mathbf{k}^\prime}= N[1-h(Q_{\mathrm{tol}}+\mu)]-fNh(Q_{\mathrm{tol}})-\log\frac{2}{\varepsilon_{\mathbf{k}^\prime}^{3}},
\end{equation}
and we can calculate the value of $\varepsilon_{\mathbf{k}^\prime}$ and then calculate the guessing probability by Eq.~(\ref{ee3}) in our main body text.

\section*{data availability}
\noindent  The data that support the findings of this study are available from the corresponding author upon reasonable request.

\section*{acknowledgement}
\noindent  We acknowledge the financial support in part by Ministration of Science and Technology of China through The National Key Research and Development Program of China grant No. 2017YFA0303901; National Natural Science Foundation of China grant No. 11474182, 11774198, 11974204 and U1738142.

\section*{author contributions}
\noindent Xiang-Bin Wang developed the theory, Jing-Tao Wang and Ji-Qian Qin contributed equally to the calculation work, Cong Jiang and Zong-Wen Yu contributed to simulation work. All authors contributed to the manuscript.

\section*{Competing interests}
\noindent The authors declare no competing interests.

\section*{Additional information}
\noindent Correspondence and requests for materials should be addressed to X.-B.W., J.-T.W. or Z.-W.Y.

\end{document}